# Coupling losses and transverse resistivity of multifilament YBCO coated superconductors

M. Polak[1], E. Usak[1,4], L. Jansak[1], E. Demencik[1], G. A. Levin[2], P. N. Barnes[2], D. Wehler[3], B. Moenter[3]

[1]Institute of Electrical Engineering, Slovak Academy of Sciences, Bratislava, Slovakia, [2]Air Force Research Laboratory, Wright-Patterson Air Force Base, USA, [3]University of Wuppertal, Germany, [4]Slovak Technical University, Faculty of Electrical Engineering and Information Technology, Bratislava, Slovakia

The corresponding author: M. Polak, elekpola@savba.sk

**Abstract.** We studied the magnetization losses of four different types of filamentary YBCO coated conductors. A 10 mm wide YBCO coated conductor was subdivided into 20 filaments by laser ablation. We measured the frequency dependence of the total losses in the frequency range 0.1 Hz $< f <$ 608 Hz. The coupling loss was obtained from the total by subtracting the hysteresis loss. The latter was measured at low frequencies since only hysteresis loss is non-negligible at frequencies below 1 Hz. The transverse resistivity, $\rho_{tr}$, was determined from the coupling losses; it was assumed that the sample length is equal to half of the twist pitch. The values of $\rho_{tr}$ deduced from the loss data were compared with those obtained by the four-point measurements with current flowing perpendicular to the filaments. Preliminary results indicate the existence of electrical contacts between the superconducting filaments and the substrate in some areas of the samples. This was also confirmed by the Hall probe mapping of the magnetic field in the vicinity of the tape. The measured transverse resistivity was close to the resistivity of the substrate (Hastelloy).

## 1. Introduction

A sufficiently ac-tolerant YBCO coated conductor can enable the fully-superconducting version for motors and generators in which both the field and armature windings are superconducting [1]. To reduce hysteresis losses in tapes several millimetres wide, subdividing of the tape into narrow parallel stripes is necessary [2,3]. The resulting conductor is a multifilamentary tape with parallel thin strips (filaments) separated by narrow gaps. The striation can be done using various technologies such as photolithography and wet etching [4] or laser micromachining [5].

The striated (filamentary) tapes become vulnerable to localized defects. This problem can be solved either by covering the filaments by a sufficiently thick normal metal layer with low resistivity metal/YBCO boundary, or by making a network of superconducting bridges, which allow the current sharing between filaments [6]. As shown by Amemyia et al. [7], in filamentary tapes prepared by laser micromachining the total losses contain also a coupling loss component due to the presence of electrically conductive path between the filaments and the metallic substrate [5].

The goal of this work is to determine the frequency dependence of losses in 3 different types of filamentary tapes with multiply connected filaments, prepared from YBCO coated conductors. We also compare the obtained results with the loss behavior of a non-striated tape and a sample prepared on LaAlO$_3$ substrate with electrically insulated filaments.

## 2. Loss components in a filamentary YBCO coated conductor

In samples of filamentary YBCO coated conductors the total losses have several components, as described in [8]. For tapes with relatively wide filaments the main loss components are the hysteresis losses, $W_{hyst}$, and the coupling losses, $W_{coupl}$. The hysteresis loss per unit length, P [W/m], in a tape consisting of $N_f$ isolated filaments, can be calculated using the formula of Brandt and Indenbom [9], if we neglect the interaction between the filaments:

$$P = 2\,f\,w_f\,I_{f,c}\,B_0\,[2B_c/B_0 \ln \cosh(B_0/B_c) - \tanh(B_0/B_c)]\,N_f, \tag{1}$$

where f is the frequency, $w_f$ is the half filament width, $I_{f,c}$ is the critical current in the filament [A/m]. $I_{f,c}$ is supposed to be constant, independent on the local flux density and on the local electric field. The consequence of this assumption is the independence of the per cycle hysteresis loss on the frequency. Herrmann et al.[10] measured the frequency independent hysteresis losses in 0.1 mm wide YBCO strips. However, numerous measurements of E(I) curves in YBCO samples showed that the slope of these curves can be described by the equation $I = I_0(E/E_o)^n$, where n can have values about 30 [11]. The reduction of the filament width below ~100 μm causes the reduction of $J_{f,c}$ and n [12]. The effect of the sample length is analyzed in [13].

## 3. Experimental

Sample W/1 has 7 filaments 0.5 mm wide on LaAlO$_3$ substrate, the width and length of the substrate are 4.8 mm and 40 mm. The striation was made by photolithography and wet etching. The following samples were prepared from coated conductor produced by the IBAD method on approximately 100 μm thick, 10 mm wide buffered Hastelloy substrate with a YBCO layer subsequently covered by a ~3μm thick Ag layer. Laser micromachining was used to striate the samples. The samples are 100 mm long and were cut from 2 longer pieces of coated conductor. The samples B/0, B/1 and B/2 were cut from a tape with the nominal critical current before striation $I_c$ = 160 A. The samples B/3 and B/4 were cut from a tape with $I_c$ in the range from 120 A to 140 A. The sample B/0 is non striated and B/1 is fully striated with 20 filaments (without superconducting bridges). Samples B/2, B/3 and B/4 are also divided into 20 filaments (approximately 0.5 mm wide each) that are interconnected by superconducting bridges, as shown in Fig.1

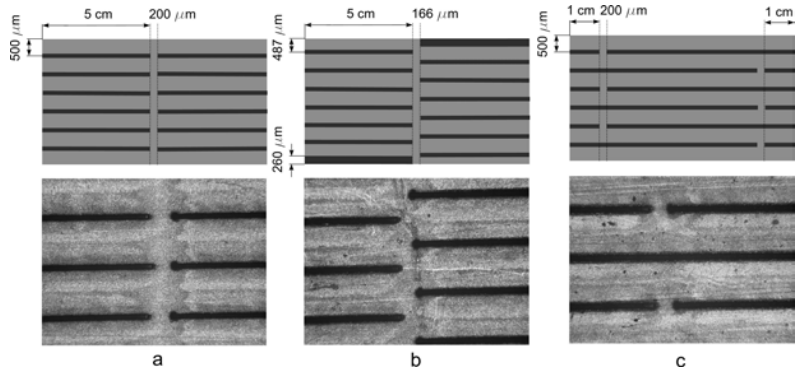

**Figure 1.** (a) – the sketch and microphotograph of sample B/2, (b) - sample B/3, and (c) -sample B /4.

The losses were measured in a copper dipole coil with the working space with diameter of 40 mm and the field/current constant of 0.97 mT/A at 77 K. At low frequencies (below 0.5 Hz) the losses were measured using pick-up coils and an analog integrator. Above 15 Hz we measured the time dependence of the pick-up coil difference voltage and the losses were determined using the first harmonics obtained by Fourier analysis of the signal.

## 4. Results and discussion

Fig. 2 shows the measured dependence of the loss per cycle in samples W/1 and W/2. They are pure hysteresis losses. The increase of $A_1$ per the change of f by a factor of 10 is ~14% and ~15% for W/1 and W/2, respectively.

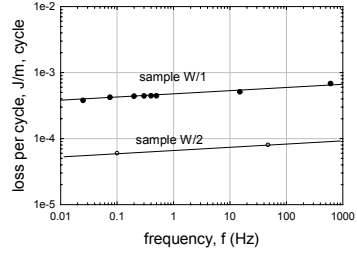
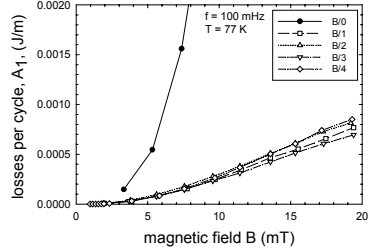
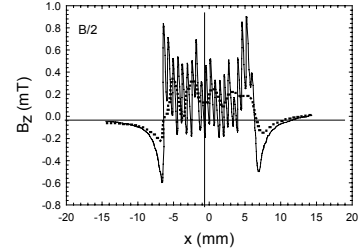

**Figure 2.** Loss per cycle $A_1$ vs. frequency for samples W/1 and W/2 at $B_m$ = 54 mT

**Figure 3.** The hysteresis losses of samples B/0, B/1, B/2, B/3 and B/4 measured at low frequency (100 mHz) and 77 K

**Figure 4.** Magnetic field in the vicinity of sample B/2 carrying magnetization currents

In Fig.3 we plotted the loss per cycle, $A_1$, measured at frequency 100 mHz, when coupling loss is negligible. As expected, the largest loss is in sample B/0, the difference of losses in the filamentary samples is relatively small. The smallest hysteresis loss at 20 mT (~7.2 x$10^{-4}$ J/m) is in sample B/3, losses in sample B/4 are larger by about 25% than those of B/3 (~9 x$10^{-4}$ J/m).

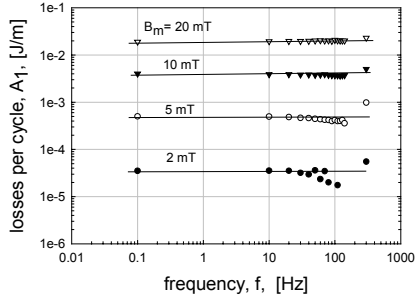
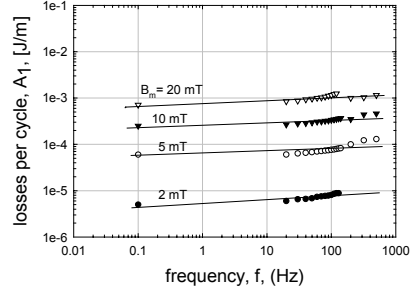
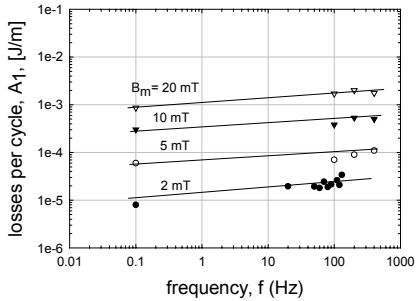
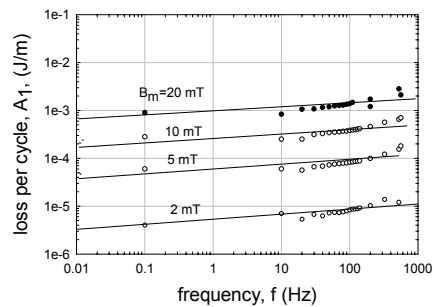

**Figure 5.** The dependence of losses per cycle on the frequency of the magnetic field at various field amplitudes: a- sample B/0, b – sample B/2, c- sample B/3, d – sample B/4

We mapped the magnetic field in the vicinity of all samples carrying magnetization currents. The measurements showed that the filament separation in some parts of samples B/2 and B/4 is not perfect, so that the effective sample width is a multiple of the filament width 0.5 mm (see Fig.4).

In Fig.5 a, b, c and d we show the losses per cycle of samples B/0, B/2, B/3 and B/4, as a function of the frequency changing from 0.1 Hz to 608 Hz. The largest loss increase has sample B/2 followed by B/4 and B/3. Sample B/0 showed very small increase of losses with increasing frequency (~5 % only). The hysteresis losses, A, are proportional to the critical current, $I_c$, which increases with increasing frequency, f, as it is controlled by the electric field E ~ f. According to the experiments with W/1 and W/2, the hysteresis losses in samples B/2, B/3 and B/4 at 500 Hz were determined by multiplying the losses determined at 0.5 Hz by the factor 3 x 12% =1.36 (the value 14 % per the decade of f measured at 54 mT was reduced due to smaller $B_m$ = 20 mT to 12 %). The coupling losses at 500 Hz, $W_{coupl}$, were determined by subtracting the hysteresis losses determined at 500 Hz from the measured total losses. The transverse resistivity $\rho_{tr}$ was calculated using the expression for the coupling losses per unit volume

$$W_{coupl} = (1/4\rho_{tr})(f B_m 2 L)^2 \qquad (2)$$

where L is the sample length and f is frequency. The values of $\rho_{tr}$ obtained for samples B/2, B/3 and B/4 are 2.5 x $10^{-6}$, 1.05 x $10^{-5}$, 3.17 x$10^{-6}$ Ωm, respectively. As seen, the values of $\rho_{tr}$ for B/2 and B/4 are close to the resistivity of the substrate (~1.3 x $10^{-6}$ Ωm).

## 5. Conclusions

The per cycle losses measured in the filamentary YBCO samples with filaments 0.1 mm and 0.5 mm wide at frequencies from ~0.1 Hz up to ~500 Hz were not frequency independent. They increased with increasing frequency by 10 to 15 % per decade of the frequency. The total loss in our samples of filamentary YBCO coated conductors on Hastelloy substrate is affected by finite interfilamentary resistance. This translates into measurable frequency dependence of per cycle loss. Substantial coupling losses were found in all filamentary samples prepared from YBCO coated conductors.

**Acknowledgements**

The financial support from AFOSR, grant number FA8655-03-1-3082 is gratefully acknowledged. We also acknowledge the support of the Center of Excellence CENG, Slovak Academy of Sciences. M.P. thanks the Alexander von Humboldt Foundation for the support of the work.